\documentclass[12pt]{article}
\usepackage{amssymb,graphicx}
\usepackage{epstopdf}
\usepackage{amsmath,amsfonts}
\usepackage{epsfig}

\def\d{\partial}
\def\l{\left(}
\def\r{\right)}

\newcommand{\be}{\begin{equation}}
\newcommand{\ee}{\end{equation}}
\newcommand{\bea}{\begin{eqnarray}}
\newcommand{\eea}{\end{eqnarray}}
\newcommand{\bg}{\begin{gather}}
\newcommand{\eg}{\end{gather}}
\newcommand{\bseq}{\begin{subequations}}
\newcommand{\eseq}{\end{subequations}}

\begin{document}
\begin{flushright}
INR-TH-2016-002
\end{flushright}
\vspace{10pt}

\begin{center}
  {\LARGE \bf More about wormholes  \\[0.3cm] in generalized 
Galileon theories} 

\vspace{20pt}

V. A. Rubakov\\
\vspace{15pt}
\textit{
Institute for Nuclear Research of
         the Russian Academy of Sciences,\\  60th October Anniversary
  Prospect, 7a, 117312 Moscow, Russia;}\\
\vspace{7pt}
\textit{
Department of Particle Physics and Cosmology,\\
Physics Faculty, M. V. Lomonosov Moscow State University\\ Vorobjevy Gory,
119991, Moscow, Russia
}

    \end{center}
    \vspace{5pt}

\begin{abstract}

We consider a class of generalized Galileon theories 
within General Relativity in
space-times of more than two spatial dimensions. We show that these
theories do not admit stable,  
static, spherically symmetric, asymptotically flat
and traversable Lorentzian wormholes.

\end{abstract}

\section{Introduction and summary}

Galileons and their 
generalizations~\cite{Horndeski:1974wa,Fairlie:1991qe,Luty:2003vm,Nicolis:2008in,Deffayet:2010zh,Deffayet:2010qz,Kobayashi:2010cm,Padilla:2012dx}  
may violate the Null 
Energy Condition (NEC) without obvious 
pathologies~\cite{Deffayet:2010qz,Kobayashi:2010cm,Genesis1,Qiu:2011cy}. 
It is therefore 
tempting to make use of these theories for constructing examples of
stable, static,
asymptotically flat, traversable
Lorentzian 
wormholes~\cite{book,Morris:1988cz,Morris:1988tu,Novikov:2007zz,Shatskiy:2008us}
in classical General Relativity,
whose putative existence necessarily requires 
NEC-violation~\cite{Morris:1988tu,hoch-visser,Hochberg:1998ha}.

The  most commonly studied
class of generalized Galileons is described by
the 
Larangians of the following 
form~\cite{Deffayet:2010qz,Kobayashi:2010cm}   (metric
signature $(+,-,\ldots ,-)$)
\be
L =   - \frac{1}{2\kappa}R + F(\pi, X) + K (\pi, X) \Box \pi  \; ,
\label{sep22-15-30}
\ee
where the first term is the Einstein--Hilbert Lagrangian,
$\kappa = 8 \pi G$, 
$\pi$ is the Galileon field,
$F$ and $K$ are arbitrary Lagrangian functions, and 
$X = \nabla_\mu \pi \nabla^\mu \pi \; , \;\;\;\;
\Box \pi =  \nabla_\mu  \nabla^\mu \pi$.
A static, sphericaly symmetric and traversable Lorentzian wormhole
in $(d+2)$-dimensional space-time is described by the
metric   
\be
ds^2 = a^2 (r) dt^2 -  dr^2 - c^2(r) \gamma_{\alpha \beta} dx^\alpha dx^\beta
\; ,
\label{jan25-16-5}
\ee
where
$x^\alpha$ and $\gamma_{\alpha \beta}$ are cordinates and
 metric on unit $d$-dimensional sphere. 
The coordinate $r$ runs from $-\infty$ to
$+\infty$, and the metric coefficients are  strictly 
positive and bounded from below:
\be
a(r) \geq a_{min} > 0 \; , \;\;\;\;\;\; c(r) \geq R_{min} > 0\; ,
\label{jan20-16-1}
\ee
where $R_{min}$ is the radius of the wormhole throat.
The wormhole geometry is assumed to be asymptotically
flat. For $d \geq 2$ (four or more space-time dimensions) this implies the
asymptotic behavior
\be
 a \to a_{\pm} \; ,   \;\;\;\;
c(r) \to \pm r \; , \;\;\;\; \mbox{as} \;\; r\to \pm \infty \; ,
\label{jan25-16-2}
\ee
where $a_{\pm}$ are positive constants.
For consistency, the 
Galileon field supposedly supporting a wormhole is also static and
spherically symmeric, $\pi = \pi (r)$.

It has been observed in Ref.~\cite{Rubakov:2015gza} that 
there is 
tension between the properties of the Galileon
energy-momentum tensor that can support a wormhole, on the one hand,
and the requirement of stability  (the absence of ghosts and 
gradient instabilities in the Galileon perturbations about the putative
solution
$\pi(r)$), on the other. In  3-dimensional space-time ($d=1$), the analysis of
Ref.~\cite{Rubakov:2015gza} was sufficient to rule out stable wormholes
 under very
mild assumption on the asymptotic
behavior of $\pi'$ at spatial infinity, cf. Ref.~\cite{Deser:1991ye}.
However, the arguments of Ref.~\cite{Rubakov:2015gza} regarding wormholes
in higher dimensional space-times were not completely conclusive.

In this paper we consider 
space-times of more than 3 dimensions ($d\geq 2$). Our purpose 
is to complete the analysis and show
that stable wormholes 
with properties \eqref{jan25-16-5}, \eqref{jan20-16-1},  \eqref{jan25-16-2}  
do not exist in theories with the Lagrangians of the
form \eqref{sep22-15-30}. The argument we present is
quite general; in particular, no assumption on the behavior of
$\pi (r)$ at $r \to \pm \infty$ is made.

The paper is organized as follows. We begin with generalities of
spherically symmetric Galileon backgrounds and perturbations about them in
Section~\ref{sec:gen}. We present our argument in Section~\ref{sec:main}
and 
conclude in Section~\ref{Discussion}.

\section{Galileon and its perturbations}
\label{sec:gen}

The Galileon energy-momentum tensor reads
\be
T_{\mu \nu} = 2 F_X \d_\mu \pi \d_\nu \pi
+  2 K_X \Box \pi \cdot \d_\mu \pi \d_\nu \pi
- \d_\mu K \d_\nu \pi - \d_\nu K \d_\mu \pi - g_{\mu \nu} F
+ g_{\mu \nu} g^{\lambda \rho} \d_\lambda K \d_\rho \pi \; ,
\nonumber
\ee
where $F_\pi = \d F/\d \pi$, $F_X = \d F/\d X$,
etc.,
and $\d_\mu K = K_\pi \d_\mu \pi + 2 K_X \nabla^\lambda \pi \nabla_\mu
\nabla_\lambda \pi$. Its components in the static case, $\pi = \pi(r)$, are
\begin{subequations}
\begin{align}
T_{0}^0 & =  -F  - K_\pi \pi^{\prime \, 2}
+ 2  \pi^{\prime \, 2} \pi^{\prime \prime}  K_X  \; ,
\label{jan3-16-10}
\\
T_{r}^r &= - 2  \pi^{\prime \, 2} F_X -F 
+ K_\pi \pi^{\prime \, 2}
+ 2  K_X \pi^{\prime \, 3} \l \frac{a'}{a} + d \frac{c'}{c} \r
\; .
\label{jan3-16-11}
\\
T^\alpha_\beta &= \delta^\alpha_\beta T^0_0
\end{align}
\end{subequations}
To derive the quadratic Lagrangian for the Galileon perturbations of
high momenta and frequencies, one writes
the full Galileon field equation 
\begin{align}
 \left( -2F_X + 2 K_\pi - 2K_{X \pi} \nabla_\mu \pi
\nabla^\mu \pi  - 
2 K_X \Box \pi \right) \Box \pi   +
\left(-4 F_{XX} + 4 K_{X \pi} \right) 
\nabla^\mu \pi \nabla^\nu \pi \nabla_\mu \nabla_\nu \pi &
\nonumber \\
 - 4 K_{XX} \nabla^\mu \pi \nabla^\nu \pi \nabla_\mu \nabla_\nu \pi
\Box \pi  +
4 K_{XX} \nabla^\nu \pi \nabla^\lambda \pi \nabla_\mu \nabla_\nu \pi
\nabla^\mu \nabla_\lambda \pi + 2 K_X \nabla^\mu \nabla^\nu \pi 
\nabla_\mu \nabla_\nu \pi &
\nonumber \\
 + 2 K_X R_{\mu \nu} \nabla^\mu \pi \nabla^\nu \pi + \ldots = 0 \; ; &
\nonumber
\end{align}
hereafter dots denote terms without second derivatives.
The last line here is a subtlety of the Galileon 
theories: the Galileon field equation involves the
second derivatives of metric, and the Einstein equations involve the
second derivatives of the Galileon~\cite{Deffayet:2010qz} (see also
ref.~\cite{Kobayashi:2010cm}), and so do the linearized equations 
for perturbations. However, the linearized theory can be reduced to the
theory of
purely Galileon perturbations. The trick is to integrate 
the metric perturbations out
of the Galileon field equation by making use of the Einstein 
equations~\cite{Deffayet:2010qz}.

 The linearized Galileon equation
can be written in the following form 
\begin{align}
 -2 [F_X  + K_X \Box \pi - K_\pi +  \nabla_\nu (K_X \nabla^\nu \pi)]
 \nabla_\mu \nabla^\mu \chi  &
\nonumber \\
-2  [2 (F_{XX} + K_{XX} \Box \pi) \nabla^\mu \pi \nabla^\nu \pi 
- 2(\nabla^\mu K_X) \nabla^\nu \pi - 2K_X \nabla^\mu \nabla^\nu \pi]
 \nabla_\mu \nabla_\nu \chi &
\nonumber\\
 + 2 K_X R_{\mu \nu}^{(1)} \nabla^\mu \pi \nabla^\nu \pi + \ldots &= 0 \; ,
\label{oct23-15-1} 
\end{align}
where $\pi$ is the background, $\chi$ is the
Galileon perturbation about this background,
and $R_{\mu \nu}^{(1)}$ is linear in metric perturbations.
We now make use of the Einstein equations $R_{\mu \nu} - 
\frac{1}{2} g_{\mu \nu}R = \kappa T_{\mu \nu}$, or
\[
R_{\mu \nu} = 
\kappa \l T_{\mu \nu} - \frac{1}{d-1} g_{\mu \nu} T^\lambda_\lambda \r \; ,
\]
linearize the energy-momentum tensor and obtain for the last term 
in eq.~\eqref{oct23-15-1}
\[
 2 K_X R_{\mu \nu}^{(1)} \nabla^\mu \pi \nabla^\nu \pi
= - 2 \kappa K_X^2 \left[-\frac{2(d-2)}{d-1} X^2 \Box \chi + 4X
\nabla^\mu \pi \nabla^\nu \pi \nabla_\mu \nabla_\nu \chi \right] + \ldots \; .
\]
The resulting linearized Galileon field equation is obtained from the
following quadratic Lagrangian:
 \begin{align*}
L^{(2)} &= [F_X  + K_X \Box \pi - K_\pi +  \nabla_\nu (K_X \nabla^\nu \pi)]
 \nabla_\mu \chi \nabla^\mu \chi
\\
&+ [2 (F_{XX} + K_{XX} \Box \pi) \nabla^\mu \pi \nabla^\nu \pi 
- 2(\nabla^\mu K_X) \nabla^\nu \pi - 2K_X \nabla^\mu \nabla^\nu \pi]
 \nabla_\mu \chi \nabla_\nu \chi
\\
& -\frac{2(d-2)}{d-1} \kappa K_X^2 X^2 \nabla_\mu \chi
\nabla^\mu \chi + 4\kappa K_X^2 X
\nabla^\mu \pi \nabla^\nu \pi \nabla_\mu \chi \nabla_\nu \chi \; .
\end{align*}
Specifying to static, spherically symmetric background, one finds
\be
L^{(2)} = a^{-2} {\cal G}^{00} \dot{\chi}^2 - 
{\cal G}^{rr} (\chi')^2 
- c^{-2} {\cal G}^\Omega \gamma^{\alpha \beta}
\d_\alpha \chi \d_\beta \chi + \dots \; ,
\nonumber
\ee
where the omitted terms do not contain derivatives of
$\chi$, and the effective metric is
\begin{align}
{\cal G}^{00} &=  F_X - K_\pi - K_X^\prime \pi'
- 2 K_X  \pi^{\prime \prime} - 
2 d K_X \frac{c^\prime}{c}\pi'  
-\frac{2(d-1)}{d} \kappa K_X^2  \pi^{\prime \, 4}\; ,
\nonumber\\
{\cal G}^\Omega &=  F_X - K_\pi - K_X^\prime \pi'
- 2 K_X  \pi^{\prime \prime} - 
2 (d-1) K_X \frac{c^\prime}{c}\pi'  - 
2  K_X \frac{a^\prime}{a}\pi' -
\frac{2(d-1)}{d} \kappa K_X^2  \pi^{\prime \, 4} \; ,
\nonumber\\
{\cal G}^{rr} &=  F_X - 2 F_{XX} \pi^{\prime \, 2}
- K_\pi 
+ K_X^\prime \pi^\prime - 2 K_X \pi^\prime 
\l \frac{a'}{a} + d \frac{c'}{c} \r
\nonumber \\
& + 2 K_{XX}  \pi^{\prime \, 2} \pi^{\prime \prime}
 + 2 K_{XX}  
\pi^{\prime \, 3} \l \frac{a'}{a} + d \frac{c'}{c} \r
+ \frac{2(d+1)}{d} \kappa 
K_X^2  \pi^{\prime \, 4}
\; .
\nonumber
\end{align}
The background is stable, provided that ${\cal G}^{00} > 0$,
${\cal G}^{rr} \geq 0$, ${\cal G}^{\Omega} \geq 0$ for every $r$, 
otherwise there are
either ghosts or gradient instabilities. We now show that the property
 ${\cal G}^{00} > 0$ cannot hold for a non-singular wormhole solution.

\section{The argument}
\label{sec:main}

A combination of the Einstein equations gives
\be
T^0_0 - T^r_r = -\frac{d}{\kappa} \frac{a}{c} \l \frac{c'}{a} \r^\prime \; .
\label{sep22-15-15}
\ee
We combine eqs.~\eqref{jan3-16-10}, \eqref{jan3-16-11} and \eqref{sep22-15-15}
to obtain the relation
\be
2 \frac{c}{a} \pi^{\prime \, 2} {\cal G}^{00}
= - \frac{d}{dr} \l 2 \frac{c}{a} K_X \pi^{\prime \, 3} +
 \frac{d}{\kappa} \frac{c^\prime}{a}  \r
- \frac{2(d-1)}{d} \kappa K_X \pi^{\prime \, 3} 
\l 2 \frac{c}{a} K_X \pi^{\prime \, 3} +
 \frac{d}{\kappa} \frac{c^\prime}{a}  \r
\label{jan3-16-12}
\ee
 It is now natural to introduce a variable
\be
{\cal Q} =  \frac{1}{c^{d-1}} \l 2\frac{c}{a} K_X \pi^{\prime \, 3} +
 \frac{d}{\kappa} \frac{c^\prime}{a} \r
\nonumber
\ee
and cast the relation~\eqref{jan3-16-12} into
\be
\frac{2}{ac^{d-2}} \pi^{\prime \, 2} {\cal G}^{00}
=  -  {\cal Q}^\prime
- \frac{d-1}{d} \kappa a c^{d -2} {\cal Q}^2
\label{jan20-16-5}
\ee
The absence of ghosts and gradient instabilities,
${\cal G}^{00} > 0$, is possible only if
\be
 \frac{{\cal Q}^\prime}{{\cal Q}^2} < - \frac{d-1}{d} \kappa 
ac^{d -2} 
\; .
\label{jan20-16-2}
\ee
We now recall that we consider the case $d \geq 2$ and that
$a(r)$ and $c(r)$ are bounded from below by positive numbers, see
eq.~\eqref{jan20-16-1}. Hence, the relation \eqref{jan20-16-2} implies
that
\[
 \frac{{\cal Q}^\prime}{{\cal Q}^2} < -{\cal C} \; ,
\]
where ${\cal C}$ is a positive constant. 

We integrate the latter relation
from $r$ to $r' > r$
and obtain
\[
{\cal Q}^{-1} (r) - {\cal Q}^{-1} (r') < - {\cal C} (r' - r) \;\;\;\;\;
\mbox{for~all}~~ r' > r \; .
\]
Suppose now that $Q(r') > 0$ at some value of $r'$. Then
\[
{\cal Q}^{-1} (r) < {\cal Q}^{-1} (r')  - {\cal C} (r' - r) \; .
\]
As $r$ decreases from $r'$ to $-\infty$, ${\cal Q}^{-1} (r)$ starts 
positive, stays bounded from above and eventually becomes bounded
by a negative number. Hence, ${\cal Q}^{-1} (r)$ crosses zero, which means that
  ${\cal Q}$ is infinite and the configuration is singular. 

${\cal Q}(r)$ cannot 
be negative everywhere either. If ${\cal Q}(r)$ is negative at some $r$, then
\[
{\cal Q}^{-1} (r') > {\cal Q}^{-1} (r)  + {\cal C} (r' - r) \; .
\]
As $r'$ increases from $r$ to $+\infty$, the right hand side increases, 
and eventually becomes positive.  ${\cal Q}^{-1} (r')$ crosses zero, and the
configuration is again singular. This completes the argument.

\section{Discussion}
\label{Discussion}

Let us end up with two remarks. First, adding conventional matter
obeying the NEC would not help: the relation \eqref{jan20-16-5} would be
modified as follows:
\be
\frac{2}{ac^{d-2}} \pi^{\prime \, 2} {\cal G}^{00}
=  -  {\cal Q}^\prime
- \frac{d-1}{d} \kappa a c^{d -2} {\cal Q}^2 -
\frac{1}{ac^{d -2}} \l T^0_0 - T_r^r \r_M \; ,
\nonumber
\ee
where  the last term is due to the conventional matter, and the NEC implies
$\l T^0_0 - T_r^r \r_M > 0$. Hence, the inequality \eqref{jan20-16-2}
would still hold, and our argument would remain valid.

Second, our argument is quite technical. It is reassuring, however, that
the generalized Galileon theories of the type \eqref{sep22-15-30} refuse
to
support wormholes, and hence time machines~\cite{Morris:1988tu}.
In this regard, it would be interesting to study whether or not
more general theories
with second-order Lagrangians and second order field equations, 
and also $p$-form 
theories~\cite{Deffayet:2010zh} have the 
same property.

The author is indebted to M. Libanov for helpful discussions
and to S.~Deser and A.~Vikman for useful correspondence.
This work has been supported by Russian Science Foundation
grant 44-22-00161.

\end{document}